
\def\foolit{\ifnum\pageno > 1 \number\pageno\fi}
\parskip 1ex
\raggedbottom

\def\ctl{\centerline}

\def\lbr{ \hfill\break }

\def\gbr{\goodbreak\noindent}

\def\ref{\par \noindent \hangindent=2pc \hangafter=1 }
\def\etal{{\it et~al.\ }}
\def\mathnew{\mathsurround=0pt}
\def\simov#1#2{\lower .5pt\vbox{\baselineskip0pt \lineskip-.5pt
	\ialign{$\mathnew#1\hfil##\hfil$\crcr#2\crcr\sim\crcr}}}
\def\simg{\mathrel{\mathpalette\simov >}}
\def\siml{\mathrel{\mathpalette\simov <}}
\def\lambdabar{\mathrel{\lower 1pt\hbox{$\mathchar'26$}\mkern-9mu
        \hbox{$\lambda$}}}
\def\frac#1#2{{#1\over#2}}
\def\Mesz{M\'esz\'aros\ }
\def\MESZ{M\'ESZ\'AROS\ }
\def\Pacz{Paczy\'nski\ }
\def\taugg{\tau_{\gamma\gamma}}
\def\eps{\epsilon}

\def\msun{{\,M_\odot}}

\def\cm{{\rm\,cm}}

\def\epm{ e^\pm }
\def\ssk{\vskip 1ex\noindent}
\def\msk{\vskip 2ex\noindent}
\def\bsk{\vskip 3ex\noindent}
\def\msun{M_{\odot}}


\def\kpc{{\rm\,kpc}}

\def\cm{~\rm{cm}}

\def\s{~\rm{s}}

\def\cmsqi{~ {\rm cm}^{-2} }
\def\cmcui{~ {\rm cm}^{-3} }

\def\MeV{~\rm{MeV}}

\def\mev{~\rm{MeV}}
\def\gev{~\rm{GeV}}

\def\erg{~\rm{ergs}}
\def\ergs{~\rm{ergs}}

\def\kpc{~\rm{kpc}}

\magnification=1100
\baselineskip=11pt
\hoffset=0.5truein
\hsize =5.5truein
\vsize =8.75truein
\parskip=0ex
\parindent 2em
\nopagenumbers
\font\Large=cmr10 at 18truept
\font\small=cmr10 at 9truept
$ $
\ssk

\ctl{\bf \Large Gamma-ray Burst Models:}
\lbr
\ctl{\bf \Large General Requirements and
Predictions$^\ast$}\footnote{}{\small$^\ast$To appear in Proc. 17th Texas Conf.
Relativistic Astrophysics, NY Acad.Sci., 1994}

\ssk\ssk\ssk

\ctl{ P. \MESZ}

\ssk\ssk

\ctl{\it Pennsylvania State University, 525 Davey Laboratory,}
\ctl{\it University Park, PA 16803, USA }

\bsk\ssk

\ctl{\bf ABSTRACT}
\msk
Whatever the ultimate energy source of gamma-ray bursts turns out to be,
the resulting sequence of physical events is likely to lead to a fairly
generic, almost unavoidable scenario: a relativistic fireball that dissipates
its energy after it has become optically thin. This is expected both for
cosmological and halo distances. Here we explore the observational motivation
of this scenario, and the consequences of the resulting models for the photon
production in different wavebands, the energetics and the time structure of
classical gamma-ray bursters.
\bsk
\ctl{\bf 1.~ OBSERVATIONAL REQUIREMENTS}
\msk
A main requirement of almost any model for gamma-ray burst sources (GRB), which
has been long appreciated, is that the ``working surface" must be
expanding due to radiation pressure. This is expected either for an extended
galactic halo ($D\siml 10^{24}\cm$) or for a cosmological ($D\sim 10^{28}\cm$)
spatial distribution (distances much smaller than $100\kpc$ are not favored by
current observational limits on the isotropy of the location distribution, e.g.
Fishman, 1994). For a characteristic observed fluence
$F=10^{-6}F_{-6}\erg\cmsqi$
the energy in a solid angle $\theta^2$ is
$$
E= 10^{43} F_{-6}D_{24}^2 \theta^2 = 10^{51} F_{-6}D_{28}^2 \theta^2 ~\erg,
\eqno(1)
$$
so the luminosity is highly super-Eddington, $E/t_b \gg 10^{38} (M/\msun)
\ergs$ for the characteristic burst durations $t_b \simg \s$ and
masses $M\sim \msun$ expected for most progenitors.
Coupled with the fact that gamma-ray photons above $\epm$ pair threshold
constitute most of the observed flux, this led to the concept of an
expanding pair-photon fireball (Cavallo and Rees, 1978, \Pacz, 1986,
Goodman, 1986). However, from the above information alone it is not
possible to determine whether the expansion will be slow or fast. If
the baryon load of the radiation-pressure ejected shells were large
($E/Mc^2 \siml 1$, the expansion would be subrelativistic, and the
flow would remain optically thick, leading to a degradation of
the gamma-rays (\Pacz, 1990).

A second element which must enter any successful GRB scenario is implied by
the detection in many GRB of photons with energies $\eps_\gamma\simg 1 \gev$.
Pair formation sets in for
$$
\eps_\gamma >{2(m_e c^2)^2}[{E_t(1-\cos\alpha)}]^{-1} \simeq
 {4 (m_e c^2)^2 } \epsilon_t^{-1} \alpha^{-2}  \eqno(2)
$$
in the laboratory frame, where $\eps_t$ is the lab frame target photon energy
and $\alpha$ is the relative angle of the two photons.
However in the same frame, causality implies $\alpha \siml \Gamma^{-1}$
where $\Gamma$ is the bulk Lorentz factor. Therefore $\Gamma^{-1}
\siml \alpha \siml 2 m_ec^2/\sqrt{\eps_t \eps_\gamma}$, or since $\eps_t \sim
1 \MeV$ is where much of the lab-frame GRB photons are,
$$
\eps_{\gamma,\mev} \siml 10^4 \eps_{t,\mev}^{-1} \Gamma_2^2 \mev ~. \eqno(3)
$$
This (e.g. Fenimore, Ho and Epstein, 1993, Harding and Baring, 1994)
is sometimes referred to as the need for a "beaming" characterized by a
factor $\Gamma$. However, it must be stressed that this refers only to relative
photon angles -- the GRB could perfectly well be emitting isotropically.
(The word ``beaming" can be misleading, and is better reserved for actual
channeling of the total emission into a restricted solid angle). In any case,
one infers a highly relativistic expansion, $\Gamma \gg 1$, and this in turn
implies that, somehow, the GRB deposits much of its radiation
energy in a low density region where it significantly exceeds the baryon rest
mass energy density. This ``low baryon loading" is {\it required} by the
observation of high energy photons $\eps_\gamma \simg 0.1-1 \gev$.

A third requirement of a satisfactory GRB model is that it must account for
the generally non-thermal appearance of the spectra, strongly suggesting
an optically thin spectrum from power-law relativistic electrons.
In principle a thermal electron distribution in a scattering-thick
medium could produce a power law up to an energy $\sim 3kT_e$, but
in order to have the photon power law extend to GeV energies the
``temperature" would have to be of the same order, and at these
energies nonthermal distributions are more likely. The requirement is
therefore very likely to be that the emitting plasma is optically thin
and nonthermal. The comoving electron density is $n'_e= L/4\pi r^2 c^3
\eta\Gamma$, where primes denote comoving-frame (CF) quantities, and
the scattering depth $\tau_e \sim n'_e \sigma_T r/\Gamma$ must satisfy
$$
\tau_e \simeq {L \sigma_T}/({4\pi r m_p c^3 \eta \Gamma^2})
\sim 1 ~ r_{12}^{-1} L_{51} \eta_2^{-3} \siml 1 ~, \eqno(4)
$$
requiring the observed radiation to be produced at radii $r_{rad}
\simg 10^{12}r_{12} L_{51}\eta_2^{-3}\cm$ (cosmological) or $r_{rad}\simg
10^7 r_7 L_{43}\eta_1^{-3}\cm$ (halo).

An additional model requirement is that, for photons observed above pair
threshold, the ``compactness parameter" (or photon-photon optical depth)
must be $\taugg < 1$ at the radius $r_{rad}$. In the lab-frame (LF) we have
$\taugg \simeq {n'_\gamma} \sigma_T r/\Gamma$, where the CF photon density in
the frame moving with $\Gamma$ is ${n'_\gamma} =L' / 4\pi r^2 c
\eps'_\gamma = L/4\pi r^2 c \eps_\gamma \Gamma$. Thus
$$
\taugg \simeq { L \sigma_T}/({4\pi r c \eps_\gamma \Gamma^2}) \siml 1
\eqno(5)
$$
for photons observed above the threshold for that $\Gamma$. Otherwise, the
above equation implies a cutoff of the photon spectrum above $\eps_\gamma$.
For the much smaller radii $r_o\sim 10^6 r_6 \cm$ expected as the initial size
of
the ``primary" GRB event from a neutron star progenitor, the compactness
parameter is large at MeV energies, which implies that the initial stages of an
impulsive fireball (or for a continuous input, the lower portion of the wind)
is
an optically thick $\epm\gamma$ flow, which eventually becomes thin at larger
radii.
\bsk
\ctl{\bf 2.~~THE GENERIC GRB MODEL}
\msk
The observational requirements discussed above provide very strong evidence
for a relativistically expanding fireball scenario. This has
some straigthforward consequences. The bulk Lorentz factor initially
grows linearly with the radius, $\Gamma \propto r$, until the plasma becomes
subrelativistic in its own rest frame. After this $\Gamma$ saturates
to the average value of the initial radiation to rest-mass ratio,
$\eta=E/Mc^2=L/{\dot M}c^2$, $\Gamma \to \eta$. The saturation radius
$r_s\sim \eta r_o$ is usually much smaller than the radiation radii $r_{rad}$
required by the (optically thin spectrum) observations.

One problem faced by simple ($\siml$ 1992) fireball models, caused by the
saturation phenomenon, is that beyond the radius $r_s$ most of the initial
energy
has been converted to kinetic energy of the baryons, the radiation energy
content
decreasing adiabaticaly. This would raise enormously the initial energy
required
to explain the observed photon luminosity (\Pacz, 1990), and in addition it
leads
to photon energy degradation. Another major problem with simple fireball models
(e.g. \Pacz, 1986, Shemi and Piran, 1990) is that the only radiation observed
would be from the fireball photons that escape at the optical thinness radius,
and these would have a quasi-thermal spectrum. In addition, if the initial
event
has a timescale comparable
to a neutron star collapse dynamic time $r_o/c\sim 10^{-3}\s$, the
observed time over which the fireball becomes thin is also comparable
(Goodman, 1986). Both the spectrum and the timescale would be unacceptable.

There are, however, at least two ways in which a relativistic fireball can
produce a longer ($10^{-1}-10^3\s$) duration, nonthermal $\gamma$-ray burst
with reasonable energy and spectrum. One of them results from the fact that the
baryons entrained in the ejecta will eventually have to run into an external
medium, and there they will reconvert their kinetic energy into radiation (e.g.
\Mesz, Laguna and Rees, 1993; Rees and \Mesz, 1992; Katz, 1994).
The external medium may either be a pre-ejected wind from the progenitor, or
the
interstellar medium. If its density is $n_o\cmcui$, deceleration occurs at a
radius
$$
r_{dec}\sim 10^{17}(E_{51}/n_0)^{1/3}(\theta \eta_2)^{-2/3}\cm \eqno(6)
$$
and the time-delayed LF duration of the burst is
$$
t_{dec} \sim \sim r_{dec}/c\Gamma^2
\sim 5\times 10^2 (E_{51}/n_0)^{1/3}\theta^{-2/3}\eta_2^{-8/3}\s.
\eqno(7)
$$
Here the total (initial) energy $E\sim 10^{51}E_{51}$ in a solid angle
$\theta^2$ is assumed to be released during an intrinsic time shorter than
$t_{dec}$ (otherwise, the intrinsic timescale $t_w$ would be the observed
duration). The total photon energy produced in the deceleration, for very
modest
subequipartition magnetic fields which ensure high radiative efficiency, is the
entire baryon kinetic energy, comparable to the initial burst radiative energy.
The large radius $r_{rad}\sim r_{dec}$ ensures optical thiness, and the strong
deceleration and reverse shocks ensure ideal conditions for relativistic
particle
acceleration leading to synchrotron and inverse Compton (IC) nonthermal
radiation.

In addition, shocks may also arise internally in the ejecta itself, before
any deceleration by the external medium occurs.
Such internal shocks could arise for various
reasons. For instance, the energy or matter input may be time-variable, so
that higher $\Gamma$ shells overtake lower $\Gamma$ shells (e.g \Pacz~and
Xu, 1994, Rees and \Mesz, 1994). If the energy or matter input at the base of
the wind occurs during an intrinsic time $t_w$ but is modulated on some
timescale
$t_v$ (shorter than $t_w$) with $\Delta \Gamma\sim \Delta \eta \sim\eta$ around
an average final Lorentz factor $\Gamma\sim \eta$, an overtaking collision
(internal dissipation shock) occurs at
$$
r_{dis} \sim c t_v \eta^2 \sim 3\times 10^{14} t_v \eta_2^2 \cm. \eqno(8)
$$
This occurs beyond the wind Thomson photosphere
$$
r_{ph}\sim {\dot M}\sigma_T /4\pi m_p c^2 \eta^2 \sim 10^{12}
L_{51}\eta_2^{-3}\cm
\eqno(9)
$$
(larger than the saturation radius) for $0.3(L_{51}/t_v)^{1/5}\siml \eta_2
\siml
10^2 (L_{51}/t_v)^{1/4}$. Also, in general $r_{dis} < r_{dec}$. For the
magnetic
fields turbulently generated in the shocks, or for a young pulsar wind, the
radiative efficiency of the internal shocks is sufficient to radiate an
appreciable fraction of the wind bulk kinetic energy.
Other mechanisms for randomizing the wind energy
might be the dissipation of Alfv\'en turbulence beyond the photosphere
(Thompson, 1994; see also Duncan and Thompson, 1992) or the conversion of
Poynting flux into photons after the charge density becomes insufficient to
sustain the frozen-in magnetic field of a young pulsar type wind (Usov, 1994,
1992). Convective Rayleigh-Taylor instabilities may also arise and become
nonlinear beyond the saturation radius (Waxmann and Piran, 1994). Below the
photosphere this would become pressure that reaccelerates the flow, while above
the photosphere it would be expected to result in freely coasting shells of
different $\eta$ as discussed above.
\bsk\gbr
\ctl{\bf 3.~~SOME OBSERVATIONAL CONSEQUENCES}
\bsk
What are some of the predictions of the dissipative relativistic fireball
scenario?

One consequence of models based on this scenario is that emission is expected
at energies other than in gamma-rays. A simultaneous burst of the same duration
and low but significant fluence is predicted at X-ray and optical energies
(\Mesz and Rees, 1993b). Breaks in the photon power-law spectrum are also
predicted (\Mesz, Rees and Papathanassiou, 1994), in qualitative agreement with
$\gamma$-ray spectra (Band, \etal, 1994), and detailed comparison of such
calculations with observations could provide useful insights into the source
physics. For a typical MeV fluence $F_\gamma\sim 10^{-6}\erg\cmsqi$ the optical
and X-ray fluences predicted for cosmological models are compatible with the
few
detected X-ray flashes as well as the lack of widespread X-ray identifications
(X-ray paucity), and with the lack (so far) of optical detections. Typically ,
however, they are above the expected HETE threshold (Ricker, \etal, 1994) of
$\sim 10^{-10}\erg\cmsqi$. As an example, for some cosmological models in the
optical U-band, $m_U\sim 11 -2.5 \log(F_u/10^{-9}\erg\cmsqi)+2.5\log(t_b/\s)$,
e.g. \Mesz, Rees and Papathanassiou, 1994.
Other satellite or ground-based observations triggered by BATSE via systems
such as BACODYNE could, for appropriate sensitivities and pointing times less
than the burst duration, also detect the GRB at other wavelenghts.
Previous attempts at simultaneous optical detection (e.g. Vanderspek, \etal,
1994, Krimm, \etal, 1994) did not yet have the sensitivity needed for a
meaningful comparison, while most other searches were not simultaneous.

A much delayed (days to weeks) radio outburst at the mJy level could also
become observable when the ejecta has expanded sufficiently for the radio
opacity to become negligible (\Pacz and Rhoades, 1993).

Simultaneous (and delayed) GeV emission is also a fairly straightforward
consequence of this scenario, for electron power-law indices not too steep
and moderate shock magnetic field strengths (which can be inferred from MeV
spectra). A sustained or delayed GeV emission (as observed, e.g. by Hurley,
\etal, 1994) can be understood in terms of a wind with internal dissipation
{\it and} an external deceleration shock (\Mesz~and Rees, 1994). The wind,
with an $\eta\siml 10^2$ and duration $t_w$ produces via internal shocks
an MeV burst extending up to $\siml$ few GeV, and also produces a longer event
($t
\sim t_{dec}$, which may be up to hours, see eq[7]) as the wind is decelerated
in the external medium. At these $\eta$ the external burst is below BATSE
threshold at MeV, but is prominent at GeV and due to the low compactness it
extends easily above 30 GeV. An alternative possibility for delayed GeV bursts
is discussed by Katz, 1994.
\bsk\gbr
\ctl{\bf 4.~~DISCUSSION AND PROSPECTS }
\msk
A dissipative relativistic fireball scenario is motivated (and practically
required) by the key observations discussed in \S 1. It is to a large degree
generic, and is expected whether the ``ultimate source" is, e.g., a young
high-field pulsar, a failed supernova Ib, a neutron star binary merger, a
halo neutron star quake, or comets crashing into magnetospheres. The basic
assumption common to all such sources is that they deliver the required
energy in a small initial volume ($r_o\siml 10^6-10^7\cm$) in a short time.
Whatever the ultimate source is, it should in any case remain hidden behind an
opaque $\epm\gamma$ veil, and the subsequent evolution of the fireball (during
which the object manifests itself observationally) is independent of the birth
details. This may be phrased as a {\it GRB ``No-Hair" theorem}: the only thing
that characterizes observationally a GRB is the initial energy, the initial
volume and (possibly) the initial energy deposition timescale. Knowing the
exact
nature of the primary GRB source would, of course, greatly help in estimating
expected rates per galaxy per year and details of the spatial distribution.
However, this should be (to a good first approximation) irrelevant for
understanding the physics of the observable intrinsic GRB properties.

A fireball model is, in principle, expected also if GRB arise in the galactic
halo.
In this case the nature, the rate of events and the spatial distribution
might be understood (see, e.g. Podsiadlowski, Rees and Ruderman, 1995; Colgate
and Leonard, 1994; Wasserman and Salpeter, 1994) if one makes some new
assumptions about the source physics.  In a cosmological setting, on the other
hand, there are plausible astrophysical sources with relatively uncomplicated
physics that could supply the observed rate and distribution (e.g. Eichler,
\etal, 1989, Narayan, \Pacz and Piran, 1992, \Mesz~and Rees, 1992, Woosley,
1993).
The dissipative fireball scenario described above explains then in a
straightforward manner (\Mesz and Rees, 1993a, \Mesz, Laguna and Rees, 1993)
the
GRB energies, the typical overall durations and the non-thermal spectra. What
is
not specifiable in detail in such a model is how the energy gets deposited in a
low density region to provide a high $\Gamma$, which is puzzling either for
halo
or cosmological sources. However, the conclusion that this {\it does} occur
seems
unavoidable, since the observations demand such conditions (see \S 1).

Several other questions can be addressed within the context of such models.
One of them is the difference in spectral and temporal properties of halo
and cosmological models. The former have smaller total energy and lower
magnetic field strengths in the wind or at the shocks. If the minimum
particle energies accelerated in the shocks were very high this could still
yield halo GRB spectra which satisfy the X-ray paucity and have acceptable
MeV breaks, but the time-delayed durations are all less than a second
(\Mesz~ and Rees, 1993b). This may be helped if the durations are explained
with a (so far unspecifiable) internal time $t_w$, but they would still tend
to overproduce X-rays. In cosmological models,
on the other hand, either the calculated dynamic time $t_{dec}$ or
an internal time $t_w$ can give acceptable spectra. Details of the
spectrum at several wavelenghts, if observed simultaneously, could
provide discriminants between these cases.

Another question that may be addressed in these models is the reported bimodal
duration distribution (Kouveliotou, \etal, 1993 and these proceedings). One
possibility is that the durations are given by the deceleration time (7) and
bursts occur in two main types of external environment. One may speculate,
e.g.,
that the progenitor's random spatial velocity causes a fraction of the bursts
to occur in the galactic halo, while others occur in the disk. The density
contrast would be at least $10^{-3}$ and the duration difference at least a
factor $\Delta n_o^{-1/3}\sim 10$. Alternatively, there may be a fraction of
bursts for which deceleration occurs in a pre-ejected denser wind,
while in others the latter is unimportant compared to the ISM.

{\it Acknowledgements:} I am grateful to Martin Rees for many discussions
and insights into these problems. The research is partially supported through
NASA NAGW-1522 and NAG5-2362.
\bsk\gbr
\ctl{\bf References}
\bsk
\ref Band, D., \etal, 1993, Ap.J., 413, 281
\ref Cavallo, G. and Rees, M.J., 1978, M.N.R.A.S., 183, 359
\ref Colgate, S.A. and Leonard, P.J.T., 1994, in {\it Gamma-ray Bursts}, ed.
  G. Fishman, \etal, p. 518 (AIP 307, NY)
\ref Duncan, R.C. and Thompson, C., 1992, Ap.J.(Letters), 392, L9
\ref Eichler, D., Livio, M., Piran, T. and Schramm, D., 1989, Nature, 340, 126
\ref Fenimore, E.E., Epstein, R.I. and Ho, C., 1993, Astr.Ap.Suppl, 97, 59.
\ref Fishman, G., these proceedings
\ref Goodman, J., 1986, Ap.J., 308, L47
\ref Harding, A.K. and Baring, M.G., 1994, in {\it Gamma-ray Bursts}, ed.
  G. Fishman, \etal, p. 520 (AIP 307, NY)
\ref Hurley, K., \etal, 1994, Nature, 372,652
\ref Katz, J.I., 1994, Ap.J., 422, 248
\ref Katz, J.I., 1994, Ap.J.(Lett.), 432, L27
\ref Kouveliotou, C., \etal, 1993, Ap.J., 413, L101
\ref \Mesz, P. and Rees, M.J., 1992, Ap.J., 397, 570
\ref \Mesz, P. and Rees, M.J., 1993a, Ap.J., 405, 278
\ref \Mesz, P. and Rees, M.J., 1993b, Ap.J. (Letters), 418, L59.
\ref \Mesz, P., Laguna, P. and Rees, M.J., 1993, Ap.J., 415, 181.
\ref \Mesz, P. and Rees, M.J., 1994, M.N.R.A.S., 269, L41
\ref \Mesz, P., Rees, M.J. and Papathanassiou, H., 1994, Ap.J., 430, L93
\ref Narayan, R., Paczynski, B. and Piran, T., 1992, Ap.J.(Letters), 395, L83
\ref Paczy\'nski, B., 1986, Ap.J.(Lett.), 308, L43
\ref Paczy\'nski, B., 1990, Ap.J., 363, 218.
\ref Paczy\'nski, B. and Rhoads, 1993, Ap.J., 418, L5
\ref Paczy\'nski, B. and Xu, G., 1994, Ap.J., 427, 708
\ref Podsiadlowski, P., Rees, M.J. and Ruderman, M., 1995, M.N.R.A.S., in press
\ref Ricker, G., \etal, 1992, in {\it Gamma-ray Bursts}, C. Ho, R. Epstein
  and E. Fenimore, eds. (Cambridge U.P.), p. 288
\ref Rees, M.J. and \Mesz, P., 1992, M.N.R.A.S., 258, 41P
\ref Rees, M.J.  and \Mesz, P., 1994, Ap.J.(Letters), 430, L93-L96
\ref Shemi, A. and Piran, T., 1990, Ap.J.(Lett.), 365, L55
\ref Thompson, C., 1994, M.N.R.A.S., 270, 480
\ref Usov, V.V., 1992, Nature, 357, 472
\ref Usov, V.V., 1994, M.N.R.A.S., 267, 1035
\ref Wasserman, I. and Salpeter, E.E., 1994, Ap.J., 433, 670
\ref Waxmann, E. and Piran, T., 1994, Ap.J., 433, L85
\ref Woosley, S., 1993, Ap.J., 405, 273
\end